\documentclass[aps,prl,floats,twocolumn,showpacs,groupedaddress]{revtex4}

\usepackage{graphicx}
\usepackage{amsfonts}
\usepackage{amsthm}

\usepackage{color}
\definecolor{Red}{rgb}{0.8,0,0}
\definecolor{Black}{rgb}{0,0,0}

\newcommand{\DeclareMathOperator}[2]{\def#1{\mathop{\mathrm{#2}}\nolimits}}

\DeclareMathOperator{\Tr}{Tr}

\DeclareMathOperator{\sgn}{sgn}

\newcommand{\ra}{\rangle}

\def\be{\begin{equation}}
\def\ee{\end{equation}}
\def\ba#1{\begin{array}{#1}}
\def\ea{\end{array}}
\def\bn{\begin{enumerate}}
\def\en{\end{enumerate}}

\def\beq{\begin{equation}}
\def\eeq{\end{equation}}

\def\vkp{{\vec{k}}_\parallel}

\begin{document}

\title{Edge modes in band topological insulators}
\author{Lukasz Fidkowski}
\affiliation{Microsoft Research, Station Q, University of California, Santa Barbara, California 93106}
\author{T. S. Jackson} 
\author{Israel Klich}
\affiliation{Department of Physics, University of Virginia, Charlottesville, Virginia 22904}

\begin{abstract}

We characterize gapless edge modes in translation invariant topological insulators.  We show that the edge mode spectrum is a continuous deformation of the spectrum of a certain gluing function defining the occupied state bundle over the Brillouin zone (BZ).  Topologically non-trivial gluing functions, corresponding to non-trivial bundles, then yield edge modes exhibiting spectral flow.  We illustrate our results for the case of chiral edge states in two dimensional Chern insulators, as well as helical  edges in quantum spin Hall states.
\end{abstract}


\maketitle

The study of topological phases of matter has been an exciting field of research since the discovery of the integer quantum Hall effect (IQHE) in the 1980's.  Recently, the discovery of the quantum spin Hall effect (QSHE) \cite{KaneMele1, HgTe0, HgTe1} and three dimensional topological insulators \cite{MooreBalents, Roy, FuKaneMele, FuKane-BiSb, BiSb} has shown that interesting physics occurs even in these simple band models.  The common feature uniting these materials is a topological ``twisting" of the band structure over the BZ; stated mathematically, the invariant is the K-theory class of the occupied state vector bundle over the BZ \cite{SRFL, kitaev-2009}.  Another striking characteristic of these materials is gapless edge modes.  In specific cases, like the IQHE or even QSHE their existence is guaranteed by various arguments \cite{laughlin, halperin,qwz, yh1,yh2} but it is natural to ask whether there is a more direct connection between bulk invariants and gapless edge modes.  
Certainly, there are cases of non-trivial topology, such as stacked IQHE planes, where edges cut parallel to the planes can be gapped.  Given that non-trivial topology by itself does not imply gapless edge modes, when do protected gapless edge modes occur?

In this Letter, we give a necessary and sufficient condition for protected gapless edge modes in terms of topological data that define the occupied state bundle.  
We study a planar edge perpendicular to a crystal direction, giving a preferred splitting of the BZ torus $T^d$ into $d-1$ ``parallel" directions and a ``perpendicular" direction: $T^d = T^{d-1} \times S^1$.  With respect to this splitting, one may 
regard the occupied state bundle as a bundle on $T^{d-1}$, extended trivially to $T^{d-1} \times [0, 2\pi]$, and then glued along the boundary $(d-1)$-tori at $0$ and $2\pi$.  This gluing is encoded in a function $U_g: T^{d-1} \rightarrow U(N)$, where $N$ is the number of occupied bands.  The non-Abelian Berry connection gives a natural way to locally straighten the fibers in the perpendicular direction, and $U_g$ can then be defined via parallel transport along the perpendicular $S^1$: \begin{equation} \label{Ug} U_g(\vkp) = \exp \left( i \int_0^{2\pi} A_\perp (\vkp, k_\perp) dk_\perp \right), \end{equation} where $A_\perp$ is the perpendicular component of the Berry connection, and the exponential is path-ordered.  The gluing function $U_g$ has a simple interpretation in terms of localized Wannier functions \cite{fkpump}: being the exponential of the perpendicular Berry-covariant derivative in momentum space, $U_g$ is, in real space, the exponential of the perpendicular position operator, projected into the occupied bands.  The corresponding eigenfunctions are localized Wannier states $|\chi_n(r-la)\rangle$, where $a$ is the unit cell spacing.  Suppressing parallel components, the $|\chi_n(r-la)\rangle$ are related to Bloch functions $|u_n(k)\rangle$ via a generalized Fourier transform \begin{equation} |\chi_n(r-la)\rangle=\frac{1}{2\pi} \int dk_\perp e^{i k_\perp(r-la)} |u_n(k)\rangle. \end{equation}  The eigenvalues of $U_g$ are then of the form $\exp ({2 \pi i \phi_n})$, where $1\leq n \leq N$ and $\phi_n$, defined modulo $1$, gives real space position of the center of the Wannier function $\chi_n$ modulo the unit cell.  Our central claim is that the spectrum $\{\phi_n(\vkp) + l\}$, where $l$ ranges over the integers, can be continuously deformed into the edge mode spectrum.  Thus the two spectra have the same topology, so spectral flow of edge modes is equivalent to spectral flow of the Wannier centers.  Because the latter is a property only of the bulk, this connection yields a criterion for the existence of protected gapless edge modes purely in terms of the bulk band structure.  After deriving the general result, we illustrate it by example for Chern insulators and QSH systems; in general, it can be applied to systems in any dimension and symmetry class.


When constructing the edge, we have a choice of boundary conditions, but all such choices are deformable to each other and hence, in the presence of a bulk gap, have the same topology.  Furthermore, in the bulk the gapped Hamiltonian can be continuously deformed to one which is {\it spectrally flat}, without closing the bulk gap and hence without changing the topology of the edge states.  Thus, we start with a spectrally flat tight binding Hamiltonian $H=1-2P$ defined on an infinite lattice ${\mathbb Z}^d$, where \begin{equation} P^{\alpha\beta}_{rs} = \int_{\text{B.Z.}} \frac{d^d k}{(2\pi)^d} \, e^{i(r-s)} P_k^{\alpha \beta}. \end{equation}  $r,s \in {\mathbb Z}^d$ are discrete spatial lattice indices, $k \in T^d$ the dual momentum, and $\alpha, \beta = 1, \ldots, N$ the band indices.  $P_k$ satisfies $P_k^2 = P_k$, and is thus a projector defining the occupied states at momentum $k$.  We choose a convenient way to introduce an edge to the system by defining \begin{equation} \label{he} H_e= P V_0(x) P + (1-P). \end{equation}  Here 
\begin{equation}
V_0(x) = \left\{ \begin{array}{ll}
1, &\text{for } x_\perp \leq 0\\
-1 &\text{for } x_\perp >0,
\end{array}\right. \end{equation}
and $x_\perp$ is the component of $x$ perpendicular to the edge.  $H_e$ has exponentially decaying hopping because $P = (1-H)/2$ does, and $V_0$ is local.  To the right, it approaches the spectrally flat form $H_{\text{right}} = 1-2P$ exponentially fast, while to the left it approaches the trivial atomic limit with negative chemical potential, $H_{\text{left}} = 1$.  Thus it models a boundary between topological insulator and vacuum.  Since $P$ and $1-P$ act in orthogonal blocks, the entire spectrum of $H_e$, aside from the trivial eigenvalue $1$, is contained in $P V_0(x) P$.

To analyze the spectrum of $P V_0(x) P$, we first note that $\vkp$, the momentum parallel to the cut, is conserved, so that we can study each $\vkp$ sector separately.  We first prove that the spectrum of $P V_0 (x) P$, in each $\vkp$ sector, is discrete.  These discrete levels correspond to edge sub-bands; in contrast to a generic gapped Hamiltonian, the special one defined in (\ref{he}) has no band continuum, but rather only accumulation points at $\pm 1$, corresponding to edge sub-bands penetrating further and further into the bulk.  Demonstrating this for fixed $\vkp$ is effectively a one dimensional problem, so to avoid cumbersome notation we simply drop $\vkp$ and assume we have a one dimensional system in the following proof.  

It is useful to view $P$ as a matrix with respect to the partition of the Hilbert space to the left and right of the cut: \begin{equation} P = \left( \begin{array}{ll} P_{LL} & P_{RL} \\ P_{LR} & P_{RR} \end{array} \right). \end{equation}  
$P_{LR},P_{RL}$ are compact operators, since their matrix elements decay exponentially, implying that $P_{LL}+P_{RR} = P-P_{LR}-P_{RL}$ has a discrete spectrum with possible accumulation points at $0,1$.  Since $P_{LL}$ and $P_{RR}$ act in orthogonal subspaces, they have discrete spectra as well, as do $H_{LL/RR} = 1-2P_{LL/RR}$.  The latter have accumulation points at $\pm 1$.  Elementary computation now gives \begin{equation} \label{fe} P V_0(x) P - (1-P) V_0(x) (1-P) = \left(\begin{array}{ll}-H_{LL} & 0\\ 0 & H_{RR} \end{array} \right) \end{equation}  Since $P V_0(x) P$ and $(1-P) V_0(x) (1-P)$ act in orthogonal subspaces, both must have discrete spectra with possible accumulation points at $\pm 1$, as desired.  We note here that the spectrum of $H_{RR}$ is directly related to the entanglement spectrum \cite{haldane-li, bernevig, nba, fidkowski, ari1, bernevig-i, ari-i, peschel, klich2006lower} via Peschel's construction \cite{peschel}.  Under mild non-degeneracy assumptions, namely the lack of bulk occupied states localized {\it exactly} on one side of the cut, the spectra of $P V_0(x) P$ and $H_{RR}$ are actually the same.

We now deform the spectrum of $P V_0(x) P$ to that of $(1/{2\pi i}) \log U_g$.  This deformation will be continuous in $\vkp$ and preserve discreteness at each fixed $\vkp$, thus preserving topological properties as well.  For convenience we again fix $\vkp$ to work with a one dimensional system in the following argument.  Recall that $(1/{2\pi}) \log U_g$ has spectrum $\{ \phi_n + l \}$, where $0 \leq \phi_n < 1$ and $l \in {\mathbb Z}$.  We first re-interpret this as the spectrum of the Berry gauge-covariant derivative $-i\nabla_B = -i\partial_{k_\perp} + A_\perp$ acting on sections of the occupied state bundle.  Indeed, defining a covariantly-constant frame $|u_n(k_\perp)\ra$ we see, using (\ref{Ug}), that $|u_n(2\pi)\ra = (U_g)^m_n |u_m(0)\ra$; we can furthermore ensure that $U_g$ is diagonal in the latter expression.  Then the eigenfunctions of $-i \nabla_B$ are localized Wannier states $\chi_n(r-la) = \int {dk_\perp} \exp {i k_\perp (\phi_n+l)} \,|u_n(k_\perp)\ra$ which have eigenvalues $\phi_n+l$, as desired.  Note also that, as an operator on Hilbert space, $-i \nabla_B = P (-i\partial_{k_\perp}) P = P x_\perp P$.  Thus, all we have to do is deform the operator $P V_0(x) P$ to $P x_\perp P$.

Define
\begin{equation}
V_t(x) = \left\{ \begin{array}{ll}
-x_\perp, &\text{for } |x_\perp| < 1/(1-t),\\[2pt]
-\sgn(x_\perp)/(1-t) &\text{for } |x_\perp| \geq 1/(1-t).
\end{array}\right.
\end{equation}
Then $V_t(x)$ interpolates between $V_0(x)$ and $x_\perp$, and $P V_t(x) P$ is the desired deformation.  Indeed, for $t<1$, $P V_t(x) P$ is a finite rank perturbation of $(1-t)^{-1}PV_{0}P$, acting only at $|x_\perp| < 1/(1-t)$, and thus cannot produce a continuous spectrum.  For $t=1$ a slight subtlety arises: $P x_\perp P$ is not a bounded operator.  To make a rigorous statement we define a bounded continuous function $h(y)$ such that $h(y) = y$ on $[-W,W]$ and $h(y) = \sgn(y) W$ for $|x| > W$, for some large $W$.  Then one can check that the spectrum of the uniformly bounded family of operators $h(P V_t P)$ is jointly continuous in $t\in[0,1]$ and $\vkp$.  This is just the statement that as one goes from $P V_0(x) P$ to $P V_1(x) P$, the spectrum evolves uniformly continuously in any finite window $[-W,W]$.

We have thus proven that the spectrum of the logarithm of $U_g$ and the edge spectrum have the same topology.  We now illustrate our result for Chern insulators and time reversal invariant systems in two dimensions.

\begin{figure}[htp]
\includegraphics[width=6cm]{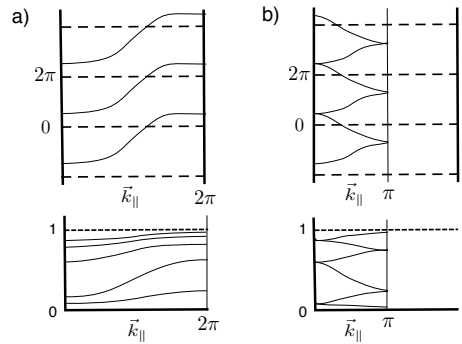}
\caption{Spectrum of gluing function (top) and entanglement - or edge - spectrum (bottom) for a) a Chern insulator with one occupied band and $\nu=1$ and b) a minimal quantum spin Hall system with nontrivial ${\mathbb Z}_2$ invariant. \label{sff1}}
\end{figure}

{\it Chern insulators:} Time reversal breaking insulators in two dimensions are characterized by an integer Chern number (TKNN invariant).  The simplest realization of nonzero Chern number has one filled band, corresponding to a line bundle over the $T^2$ BZ.  The transition function discussed above is a map from $S^1 \subset T^2$ to $U(1)$, and the element of $\pi_1(S^1, U(1))={\mathbb Z}$ which it defines is equal to $\nu$.  The spectrum of its logarithm has $\nu$ protected modes spiraling up, just as the edge spectrum, consistent with our picture - see figure \ref{sff1}.

More quantitatively, we can see spectral flow directly by evaluating \cite{kitaev-anyons} \begin{equation} \nu = 2 \pi i \Tr \left(P\left[\left[P,\Pi_x\right], \left[P, \Pi_y \right]\right]\right),\label{nu} \end{equation} where now $\Pi_{x,y}$ are projections onto the positive $x$ and $y$ half-planes respectively.  The topological nature of $\nu$ gives us a lot of freedom to deform the expression (\ref{nu}).  In particular, with translation symmetry one can deform the $\Pi_{x,y}$ into linear functions $X$, $Y$ (replacing sums with integrals), obtaining the usual expression for $\nu$ as the integral of the Berry curvature over the BZ \cite{kitaev-anyons}.  Now take an entanglement cut along $y$ (so that $x=x_\perp$ and $y=x_\parallel$).  Then it is most useful to deform only $\Pi_y$: \begin{equation} \nu =  \int dk_y \Tr C(k_y) \label{me} \end{equation} where $C(k_y) = P_{k_y} [ [P_{k_y}, \Pi_x], \partial P_{k_y} / \partial k_y]$.  We evaluate (\ref{me}) in the basis $|\psi_n(k_y)\rangle$ of eigenstates of $P_{k_y} V_0(x) P_{k_y}$: \begin{equation} P_{k_y} V_0(x) P_{k_y} |\psi_n\rangle = \lambda_n |\psi_n\rangle. \end{equation}  Using (\ref{fe}) and the fact that $\frac{\partial P_{k_y}}{\partial k_y} | \psi_n \rangle = \frac{\partial}{\partial k_y} | \psi_n \rangle$ is orthogonal to $|\psi_n \rangle$ we obtain 
\begin{eqnarray} \langle \psi_n | C(k_y) | \psi_n \rangle = \frac{\partial}{\partial k_y} \,\langle \psi_n | \Pi_x | \psi_n \rangle \nonumber = \frac{\partial}{\partial k_y} \left( \frac{1+\lambda_n}{2} \right). \end{eqnarray}  Thus (\ref{me}) is the integral of a total derivative, resulting in the expression $\nu = \sum_n ( \lambda_n(2\pi) - \lambda_n(0))/2$.  Now, $\lambda_n(2\pi) = \lambda_{n+k}(0)$ for some $k$ which is easily seen to be independent of $n$ (the case of degenerate $\lambda_n$ must be handled carefully, but all level crossings can be appropriately resolved).  $k$ is simply the number of edge modes that have to cross any given Fermi level.  The sum then telescopes and we obtain $\nu=k$, as desired.

 {\it Continuum Chern insulators:} \cite{rs, klich_cont}
As a check of the above results, one would like to have a direct calculation of the spectrum of $H_{RR}=1-2P_{RR}$.  This is tedious for a lattice Chern insulator, but easily doable in the continuum, i.e. for the IQHE.  One problem is that the continuum BZ is not a finite torus; nevertheless, we will satisfy ourselves with demonstrating spectral flow as $\vkp$ ranges over the momentum scale corresponding to the inverse magnetic length $\ell^{-1}$.  We will work on a semi-infinite cylinder of large radius; this only introduces a fine discretization of the domain $\vkp$ of the plots.

The bulk Hamiltonian $H=1-2P$ gives energy $-1$ to the $\nu$ lowest occupied Landau levels, and energy $1$ to all higher levels.  Its eigenvalues are labeled by $k_\parallel$ and a Landau level number, and the eigenstates are localized harmonic oscillator functions in the $x_\perp$ direction.  The truncated Hamiltonian $H_{RR}$, written in this basis, contains Landau level mixing terms.  Letting $n,m=0,\ldots, \nu-1$ denote occupied Landau levels, $(H_{RR})^n_m$ is equal to \begin{equation} \delta_{nm}- {2\over \sqrt{2^{n+m}  \pi n!  m! }} \int_{-\infty}^{u_{0}} du \, e^{u^2} \left( \partial^n_u e^{-u^2} \right) \left( \partial^m_u e^{-u^2} \right). \end{equation}  Here $u_0 = k_\parallel \ell$ and $u=x_\perp / \ell + u_0$.  For example

\begin{figure}[htp]
\includegraphics[width=4.5cm]{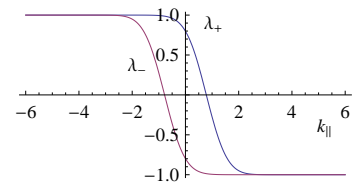}
\caption{Entanglement spectrum flow Eq. (\ref{IQH2}) for  $\nu=2$ Landau levels. \label{sff2}}
\end{figure}

 For example, for $\nu =2$ the eigenvalues of $(H_{RR})^n_m$ are plotted in Fig. \ref{sff2}; we see that both branches cross any given value of $\lambda$, displaying spectral flow.

{\it Quantum spin Hall systems:}
QSH insulators are characterized by a nonzero ${\mathbb Z}_2$ invariant.  With a choice of gauge one can express it in terms of Pfaffians \cite{KaneMele1}, or as an obstruction to trivializing the occupied state bundle in a way compatible with $T$ \cite{fkpump}.  These are equivalent to the gluing function constructed above being non-trivial.  Indeed, since the spectrum of the gluing function is given by the positions of the Wannier centers, a non-trivial gluing function is equivalent to Kramers pair-switching spectral flow of the time reversal polarization, an equivalent formulation of the $\mathbb{Z}_2$ invariant \cite{fkpump}.  

Because $T$ symmetry relates $H(\vec{k})$ to $H(-\vec{k})$, the spectrum of $U_g(k_\parallel)$ is the same as that of $U_g(-k_\parallel)$, and because $T^2=-1$ requires an even number of bands, the gluing function can be thought of as a map $U_g: [0,\pi] \rightarrow U(2N)$.  The extra constraint $T U_g T^{-1} = U_g^{-1}$ at the $T$ symmetric endpoints $k_\parallel = 0, \pi$ forces the spectra at those points to form Kramers pairs.  The non-trivial class now switches elements in these Kramers pairs as one moves from $0$ to $\pi$ (formally the classification is given by the relative homotopy group $\pi_1 (U(2N),U(N)) = {\mathbb Z}_2$).  

The key point now is that the deformation between the spectrum of $\frac{1}{2\pi i} \log U_g$ and the edge spectrum respects $T$ throughout, so Kramers partner-switching in the edge spectrum is an equally good criterion for QSH topological order.  The simplest example consists of two time-reversed filled bands, with the two eigenvalues of the gluing function splitting up as one moves away from $k_y=0$, traversing by $\pi$ in opposite directions and rejoining at $k_y=\pi$; the helical edge spectrum shares this property, consistent with our result (Fig. \ref{sff1}).

{\it Generalizations:} 

We related only one particular gluing function to edge modes.  By considering the other $d-1$ possible cuts, we gain more information.  In particular, if the edge modes for all $d$ cuts are gapped, our construction shows that the insulator is topologically trivial, i.e. can be adiabatically connected to a trivial insulator in the atomic limit.  Furthermore, we can perform similar arguments in other dimensions and symmetry classes, e.g. $3$d topological insulators \cite{rv}.

Another generalization concerns our formula for the Chern invariant in terms of the entanglement spectrum.  We can in fact apply it to {\it any} mixed state $\rho$ that is translationally invariant in $y$.  What does such an invariant represent?\

To answer this we construct the {\it purification} of our density matrix. 
Consider $\rho=Z^{-1}e^{-\sum H^{\alpha\beta}_{rr'}\Psi_{\alpha r}^\dag \Psi_{\beta r'}}$. We now double the number of sites and consider the Hamiltonian 
 \begin{eqnarray}&H_{AW}=\sum_{r\alpha\, r'\beta} (n-{1\over 2})^{\alpha\beta}_{rr'}\Psi_{\alpha r}^\dag \Psi_{\beta r'}+({1\over 2}-n)^{\alpha\beta}_{rr'}\tilde{\Psi}_{\alpha r}^\dag \tilde{\Psi}_{\beta r'}\\ \nonumber &+ \sqrt{n(1-n)}^{\alpha\beta}_{rr'}\,\tilde{\Psi}_{\alpha r}^\dag{\Psi}_{\beta r'}+\sqrt{n(1-n)}^{\alpha\beta}_{rr'}\,{\Psi}_{\alpha r}^\dag \tilde{\Psi}_{\beta r'}, \end{eqnarray} where $n={1\over 1+e^{H}}$ is the Fermi-Dirac operator and  $\tilde{\Psi}$ are the fermi operators on the auxiliary system.
Writing in block form the single particle Hamiltonian of $H_{AW}$, it is:
\[h_{WA}=
\left(
\begin{array}{ccc}
 n-{1\over 2} & \sqrt{n(1-n)}    \\
\sqrt{n(1-n)}  & {1\over 2}-n
\end{array}
\right)
\]
It is easy to check that $h_{WA}$ is of the form $P-{1\over 2}$, $P$ a projector. Hence, $H_{AW}$ is spectrally flat, describing two bands of energies $-1/2$ and $1/2$.  The ground state is the purification of our state $\rho$, also known as the ``Araki-Wyss'' representation \cite{araki1964representations}.  For a system with decaying two point function, one can check that $n$ has decaying matrix elements. Thus the $H_{AW}$ above describes a bilayer system coupled in the bulk, rather than through an edge. Indeed for a generic thermal state the entanglement entropy associated with $H$ is extensive.  However if $H$ is coming from tracing half of a system at zero temperature, the coupling terms, proportional to  $\sqrt{n(1-n)}$ will vanish exponentially inside the bulk, since $n(1-n)=0$ whenever $n=0$ or $n=1$, and we are left with a bilayer which is only coupled close to the boundary.
Thus, the entanglement Chern number represents the Chern invariant of the associated bilayer Araki-Wyss system. 

{\it Conclusions}: We have shown that the edge modes of a topological insulator are a continuous deformation of the spectrum of a gluing function defining the occupied state bundle over the BZ, giving a necessary and sufficient condition for protected gapless edge modes.


\acknowledgments  We thank B. A. Bernevig, M. Freedman, and especially A. Turner for useful discussions. IK acknowledges financial support from NSF grant No. DMR-0956053.

\bibliographystyle{apsrev} 
\bibliography{specflowrefs}

\end{document}